\documentclass{article}

\usepackage{amssymb}
\usepackage{amsmath}
\usepackage{combelow}
\usepackage{graphicx}

\def\omegat{{\widetilde{\omega}}}

\title{Rigidly-rotating quantum thermal states in bounded systems}

\author{Victor E. Ambru\cb{s}$^*$\\
{\it \small Department of Physics, West University of Timi\cb{s}oara,}\\
{\it \small Bd. Vasile P\^arvan No. 4, Timi\cb{s}oara, 300223, Romania}\\
{\it \small $^*$E-mail: victor.ambrus@e-uvt.ro}
}

\begin{document}

\maketitle

\begin{abstract}
We consider rigidly-rotating thermal states of a massless 
Klein-Gordon field enclosed within a cylindrical boundary,
where Robin boundary conditions (RBCs) are imposed. The 
connection between the parameter of the RBCs and the energy 
density and four-velocity expressed in the Landau frame is revealed. \\

\end{abstract}

\section{Introduction}

In quantum field theory, the boundary conditions (b.c.s) are imposed at 
the level of the field operator $\Phi$, or equivalently, of the quantum 
modes. The interplay between the b.c. formulation and the ensuing 
operator expectation values in various states is far from 
obvious. In this paper, we consider the connection between the 
choice of b.c.s and the thermal expectation value (t.e.v.) of 
the stress-energy tensor (SET) operator in rigidly-rotating finite 
temperature states of the massless Klein-Gordon (KG) field. We show 
that the free parameter $\Psi$ in the Robin b.c.s (RBCs) can be 
related to the values of the Landau frame macroscopic four-velocity 
and energy density on the boundary. 

The outline of this paper is as follows. In Sec.~\ref{sec:modes},
the mode solutions of the KG equation inside a cylinder are reviewed.
The procedure for constructing t.e.v.s
in rigidly-rotating systems is summarised in Sec.~\ref{sec:tevs}. 
The analysis of the SET using the Landau 
frame decomposition and the connection between the Landau velocity $v$
and $\Psi$ is presented in Sec.~\ref{sec:landau}. 
Section~\ref{sec:conc} concludes this paper.

\section{Rigidly-rotating thermal expectation values}\label{sec:modes}

Let $\hat{\Phi}(x)$ be the field operator for a massless, neutral (real) scalar field 
which is confined within a cylinder of radius $R$, obeying
the KG equation:
\begin{equation}
 \Box \hat{\Phi}(x) = 0.\label{eq:kg}
\end{equation}
The mode solutions of Eq.~\eqref{eq:kg} can be obtained as 
follows:\cite{duffy03}
\begin{equation}
 f_{j} = \frac{N_j}{\sqrt{8\pi^2 \omega_j}} 
 e^{-i\omega_j t + ik_jz + im_j\varphi} 
 J_{m_j}(q_j\rho),\label{eq:fj}
\end{equation}
where $(\rho, \varphi, z)$ are the usual cylindrical coordinates,
while $\omega_j > 0$, $k_j$ and $m_j$ are the eigenvalues of the 
Hamiltonian $\hat{H}$, longitudinal momentum $\hat{P}_z$ and $z$ component 
of the angular momentum, $\hat{L}_z$. 
In order to fix the normalisation constant $N_j$, we evaluate 
the KG inner product for $f_j$ and $f_{j'}$:
\begin{eqnarray}
 \langle f_j, f_{j'} \rangle &=& 
 i \int_{-\infty}^\infty dz \int_0^\infty \rho\, d\rho \int_0^{2\pi} d\varphi
 (f_j^* \partial_t f_{j'} - f_{j'} \partial_t f_j^*)\nonumber\\
 &=& \frac{N_j^* N_{j'} (\omega_j + \omega_{j'})}
 {2\sqrt{\omega_j \omega_{j'}}} e^{i(\omega_j - \omega_{j'}) t} 
 \delta_{m_j, m_{j'}} \delta(k_j - k_{j'}) 
 \frac{R}{q_j^2 - q_{j'}^2} \nonumber\\
 & &\times \left[ 
 J_{m_j}(q_j R) q_{j'} J_{m_j}'(q_{j'}R) - J_{m_j} (q_{j'}R) q_j J_{m_j}'(q_j R)\right],
 \label{eq:scprod_aux}
\end{eqnarray}
where a standard identity involving integrals of Bessel functions 
was employed.\cite{olver10}
Orthogonality is ensured when the transverse momenta 
$q_j \rightarrow q_{m, \ell}$ are discretised 
according to the Robin boundary conditions:\cite{romeo01}
\begin{equation}
 q_{m,\ell} R J'_m(q_{m,\ell} R) + \Psi J_{m}(q_{m,\ell} R) = 0,
 \label{eq:robin}
\end{equation}
where $\ell = 1, 2, \dots$ indexes the non-negative solutions of 
Eq.~\eqref{eq:robin} for fixed $m$ in ascending order, while 
$\Psi$ is considered to be a constant, real number.
It is easy to see that $\Psi = 0$ corresponds 
to the von Neumann b.c.s
[$J'_{m}(q_{m,\ell} R) = 0$], while the Dirichlet b.c.s 
[$J_m(q_{m,\ell} R) = 0$] 
can be recovered in the limit $\Psi \rightarrow \infty$.
Imposing $\langle f_{km\ell}, f_{k'm'\ell'} \rangle = 
\delta(k - k') \delta_{m,m'} \delta_{\ell,\ell'}$ yields:\cite{romeo01}
\begin{equation}
 N_{k,m,\ell} 
 = \frac{q_{m,\ell} \sqrt{2}}{|J_m(q_{m,\ell} R)| 
 \sqrt{q_{m,\ell}^2 R^2 + \Psi^2 - m^2}}.
 \label{eq:norm}
\end{equation}

The canonical expansion of the field operator with respect to the 
modes $f_j$ is:
\begin{equation}
 \hat{\Phi}(x) = \sum_j \left[f_j(x) \hat{a}_j + 
 f_j^*(x) \hat{a}^\dagger_j\right], \label{eq:Phi}
\end{equation}
where the one-particle creation ($\hat{a}_j^\dagger$) and annihilation 
($\hat{a}_j$) operators obey the standard commutation relation 
$[\hat{a}_j, \hat{a}^\dagger_{j'}] = \delta(j,j')$.

\section{Rigidly-rotating thermal states}\label{sec:tevs}

We now consider rigidly-rotating thermal states, corresponding to an 
inverse temperature $\beta$ and an angular velocity $\Omega$.
The thermal expectation value (t.e.v.) of an operator $\hat{A}$ 
is computed using the density operator $\hat{\rho}$ as follows:
\begin{equation}
 \langle \hat{A} \rangle_\beta = Z^{-1} {\rm tr}(\hat{\rho} \hat{A}), \qquad 
 \hat{\rho} = {\rm exp}\left[-\beta (\hat{H} + \Omega \hat{L}_z)\right],
\end{equation}
where $Z = {\rm tr}(\hat{\rho})$ is the partition function.
It can be shown that:\cite{vilenkin80}
\begin{equation}
 \langle \hat{a}_j^\dagger \hat{a}_{j'} \rangle_\beta = 
 \frac{\delta(j,j')}{e^{\beta\omegat_j} - 1}, 
 \qquad \omegat_j = \omega_j - \Omega m_j.
 \label{eq:bblocks}
\end{equation}
Equation~\eqref{eq:bblocks} is not valid when the 
co-rotating energy $\omegat_j < 0$, since in this case, the vacuum 
limit (corresponding to $\beta \rightarrow \infty$)
yields a non-vanishing value.\cite{ambrus14plb}
Moreover, modes with $\omegat < 0$ make infinite contributions 
to rigidly-rotating t.e.v.s.\cite{duffy03}
It is noteworthy that finite quantum corrections can 
still be computed perturbatively.\cite{becattini15}
It is reasonable to expect that 
t.e.v.s should stay finite for all values of $\Omega$ provided that
$\Omega R < 1$, which requires that $q_{m,\ell} R \ge m$. 
This property can be ensured only when $\Psi \ge 0$, thus 
we do not consider negative values of $\Psi$ in this paper.

Starting from the following expressions for the SET operator:\cite{callan70,groves02}
\begin{equation}
 \hat{T}_{\mu\nu} = \frac{2}{3} \nabla_{(\mu} \hat{\Phi} \nabla_{\nu)} \hat{\Phi} - 
 \frac{1}{3} \hat{\Phi} \nabla_{(\mu} \nabla_{\nu)} \hat{\Phi} - \frac{1}{6} g_{\mu\nu} 
 [(\nabla \hat{\Phi})^2 + \mu^2 \hat{\Phi}^2],\label{eq:SET}
\end{equation}
the t.e.v. of the components of the SET can be obtained using the 
mode expansion \eqref{eq:Phi} of the field operator. It is 
convenient to express the results with respect to the tetrad 
comprised of the vectors $e_{\hat{t}} = \partial_t$, 
$e_{\hat{\rho}} = \partial_\rho$, 
$e_{\hat{\varphi}} = \rho^{-1} \partial_\varphi$ and 
$e_{\hat{z}}= \partial_z$. Using the 
notation $T_{\hat{\alpha}\hat{\gamma}} \equiv 
\langle :\hat{T}_{\hat{\alpha}\hat{\gamma}}: \rangle_\beta$, the following 
results can be obtained:\cite{ambrus17plb}
\begin{eqnarray}
 T_{\hat{\alpha}\hat{\gamma}} &=& 
 \sum_{m = -\infty}^\infty \sum_{\ell = 1}^\infty 
 \int_{-\infty}^\infty \frac{N_{km\ell}^2 dk}{12\pi^2 \omega_{km\ell}
 (e^{\beta \omegat_{km\ell}}-1)} F_{\hat{\alpha}\hat{\gamma}},\nonumber\\
 F_{\hat{t}\hat{t}} &=& \left(6\omega^2 + \rho^{-2} m^2 - q^2\right)J_m^2 + q^2 J_m'{}^2,
 \nonumber\\
 F_{\hat{\varphi}\hat{t}} &=& - 6 \omega \rho^{-1} m J_m^2,\nonumber\\ 
 F_{\hat{\rho}\hat{\rho}} &=& \left(-3\rho^{-2} m^2 + 3 q^2\right) J_m^2 + 
 2 q \rho^{-1} J_m J_m' + 3 q^2 J_m'{}^2,\nonumber\\
 F_{\hat{\varphi}\hat{\varphi}} &=& \left(5\rho^{-2} m^2 + q^2\right) J_m^2 - 
 2 q \rho^{-1} J_m J_m' - q^2 J_m'{}^2,\nonumber\\
 F_{\hat{z}\hat{z}} &=& \left(6k^2 - \rho^{-2} m^2 + q^2\right) J_m^2 - 
 q^2 J_m'{}^2,\label{eq:SET_tev}
\end{eqnarray}
where it is understood that $\omega \equiv \omega_{m,\ell}$ and 
$q \equiv q_{m,\ell}$, while the Bessel functions and their 
derivatives take the argument $q_{m,\ell} \rho$.
It can be shown that the components of the SET not displayed above 
vanish for all values of $\rho$, $\beta$ and $\Omega$.

\section{Landau decomposition}\label{sec:landau}


The matrix structure of the SET given in Eq.~\eqref{eq:SET_tev} can be summarised 
as follows:
\begin{equation}
 T_{\hat{\alpha}\hat{\gamma}} = 
 \begin{pmatrix}
  T_{\hat{t}\hat{t}} & 0 & T_{\hat{t}\hat{\varphi}} & 0\\
  0 & T_{\hat{\rho}\hat{\rho}} & 0 & 0\\
  T_{\hat{t}\hat{\varphi}} & 0 & T_{\hat{\varphi}\hat{\varphi}} & 0\\
  0 & 0 & 0 & T_{\hat{z}\hat{z}}
 \end{pmatrix}.
\end{equation}
The energy density $E$ and macroscopic four-velocity $u^{\hat{\alpha}}$ can be
obtained in the Landau frame by solving the eigenvalue 
equation $T^{\hat{\alpha}}{}_{\hat{\gamma}} u^{\hat{\gamma}} = 
-E u^{\hat{\alpha}}$.\cite{landau87,rezzolla13}
The physically relevant solution for $E$ reads:
\begin{equation}
 E = \frac{1}{2}\left[T_{\hat{t}\hat{t}} - 
 T_{\hat{\varphi}\hat{\varphi}} + 
 \sqrt{(T_{\hat{t}\hat{t}} + T_{\hat{\varphi}\hat{\varphi}})^2 - 4
 T_{\hat{t}\hat{\varphi}}^2}\right],
\end{equation}
while the Landau velocity $u^{\hat{\alpha}} = \Gamma (1, 0, v, 0)^T$ 
can be characterised via:
\begin{equation}
 v = -\frac{T_{\hat{t}\hat{\varphi}}}{E + T_{\hat{\varphi}\hat{\varphi}}},\qquad 
 \Gamma = \frac{1}{\sqrt{1 - v^2}}.\label{eq:vaux}
\end{equation}
Further manipulation of the above relations gives:
\begin{equation}
 \frac{v}{1 + v^2} = -\frac{T_{\hat{t}\hat{\varphi}}}
 {T_{\hat{t}\hat{t}} + T_{\hat{\varphi}\hat{\varphi}}}.
 \label{eq:v}
\end{equation}

When Dirichlet b.c.s are employed, it is easy to see that 
$v$ vanishes on the boundary. For finite values of $\Psi$, 
$v$ is in general non-vanishing on the boundary. 
Let $v_{\rm b}$ denote the value of $v$ on the boundary. 
We now ask what is the value of $\Psi$ which ensures $v(R) = v_{\rm b}$.
Inverting Eq.~\eqref{eq:v} in order 
to obtain $\Psi$ as a function of $v_{\rm b}$ does not seem feasible. Instead, 
an iterative procedure can be established which allows the value of $\Psi$ to 
be computed numerically. Starting from:
\begin{eqnarray}
 T_{\hat{t} \hat{\varphi}}\rfloor_R &=& -
 \sum_{m,\ell} \left(1 + \frac{\Psi^2 - m^2}{q^2 R^2}\right)^{-1}
 \int_0^\infty dk \frac{2 m}{\pi^2 R^3(e^{\beta \omegat} - 1)},\nonumber\\
 (T_{\hat{t}\hat{t}} + T_{\hat{\varphi}\hat{\varphi}})_R &=& 
 \sum_{m,\ell}
 \left(1 + \frac{\Psi^2 - m^2}{q^2 R^2}\right)^{-1} \int_0^\infty dk
 \frac{2 (m^2 + \omega^2 R^2 + \frac{\Psi}{3})}
 {\pi^2 R^4 \omega(e^{\beta\omegat} - 1)},
\end{eqnarray}
where the sums over $m$ and $\ell$ run between $-\infty$ and $\infty$ 
and $1$ and $\infty$, respectively, 
it can be seen that $\Psi$ can be isolated from the last term of the second equality above:
\begin{equation}
 \Psi = \frac{\displaystyle 3 \sum_{m,\ell}
 \left(1 + \frac{\Psi^2 - m^2}{q^2 R^2} \right)^{-1} \int_0^\infty 
 \frac{dk}{\omega (e^{\beta\omegat} - 1)} \left[ 
 m \omega R - \frac{v_{\rm b}(\omega^2 R^2 + m^2)}{1 + v_{\rm b}^2}\right]}
 {\displaystyle \frac{v_{\rm b}}{1 + v_{\rm b}^2} \sum_{m,\ell} 
 \left(1 + \frac{\Psi^2 - m^2}{q^2 R^2} \right)^{-1} 
 \int_0^\infty \frac{dk}{e^{\beta\omegat} - 1}}.
 \label{eq:Psi}
\end{equation}

\begin{figure}[t]
\begin{center}
\begin{tabular}{cc}
\includegraphics[width=0.45\columnwidth]{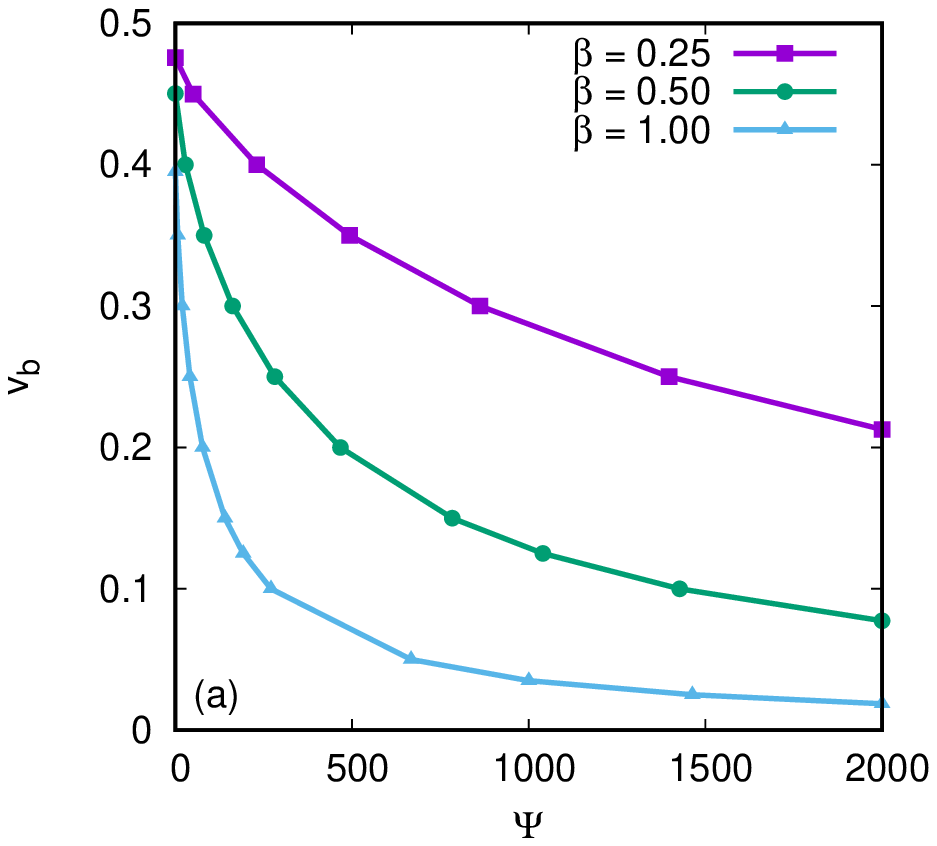} &
\includegraphics[width=0.45\columnwidth]{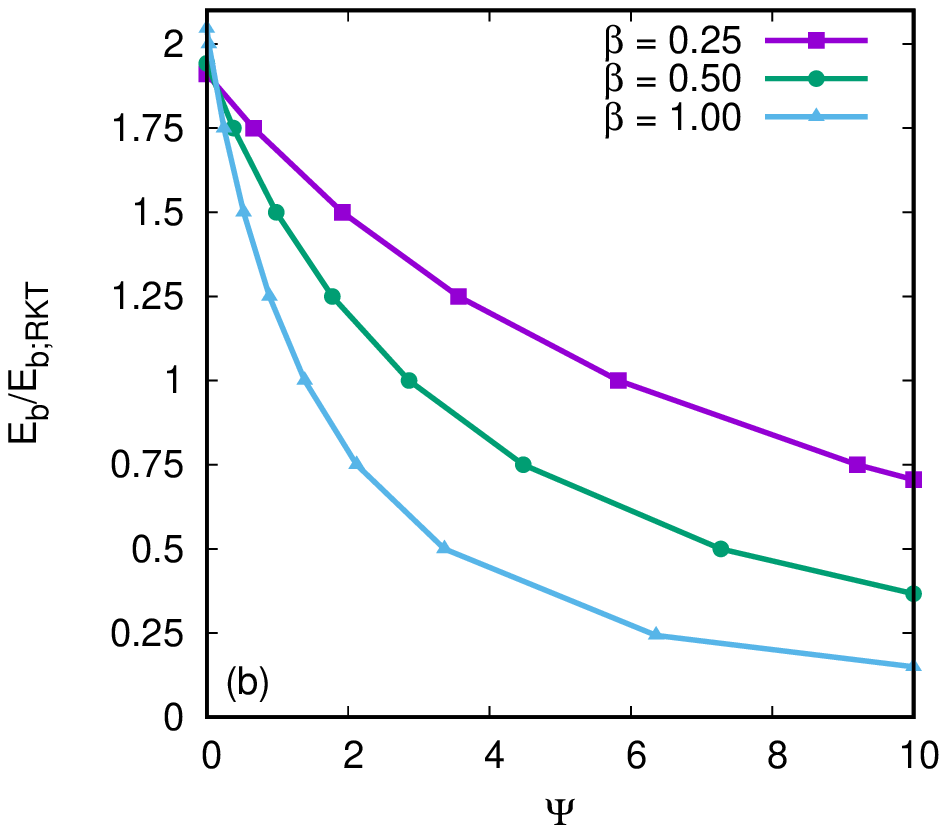}
\end{tabular}
\end{center}
\caption{The dependence of (a) $v_{\rm b}$ and (b) $E_{\rm b} / E_{{\rm b}; {\rm RKT}}$
on $\Psi$ for various values of the inverse temperature $\beta$ when 
$\Omega = 0.5$ and $R = 1$.
\label{fig:valPsi}}
\end{figure}

Equation~\eqref{eq:Psi} is solved iteratively. The value $\Psi^{(n)}$ corresponding 
to iteration $n$ is obtained by evaluating the right hand side of Eq.~\eqref{eq:Psi}
after replacing $\Psi$ with the value $\Psi^{(n-1)}$ obtained 
at iteration $n -1$, while $v_{\rm b}$ is kept fixed at the desired value. 
Starting from $\Psi^{(0)} = 0$ yields the 
convergence value within a relatively small number of iterations and the 
process seems to be stable as long as $v_{\rm b}$
can be obtained using $\Psi \ge 0$. To illustrate the 
procedure, we consider a system with $R = 1$ and $\Omega = 0.5$.
Figure~\ref{fig:valPsi}(a) shows the variation of $v_{\rm b}$ with $\Psi$ 
for $\beta \in \{0.25, 0.5, 1\}$.

It is natural to consider the relation between the Landau energy density $E_{\rm b}$ 
measured on the boundary and the energy density $E_{{\rm b}; {\rm RKT}}$ expected
for a rigidly-rotating Bose-Einstein gas, for which\cite{ambrus16prd2}
\begin{equation}
 v_{\rm RKT} = \rho \Omega, \qquad 
 E_{\rm RKT} = \frac{\pi^2 \Gamma^4_{\rm RKT}}{30\beta^4}.\label{eq:RKT}
\end{equation}
An iterative scheme for finding $\Psi$ for a prescribed value of 
$E_{\rm b}$ involves working with quadratic functions with 
respect to the SET components. The stability and efficiency of such a 
scheme is questionable. Instead, we employ a bisection algorithm to 
find the value of $\Psi$ corresponding to $E_{\rm b}$. 
Typically, the ratio $E_{\rm b} / E_{{\rm b};{\rm RKT}}$ 
characterising the departure of the quantum state from 
the expected rigid-rotation profile ranges from 
$\sim 0.1$ for Dirichlet b.c.s to $\sim 2$ for von Neumann b.c.s. 
The dependence of $E_{\rm b} / E_{{\rm b};{\rm RKT}}$ on $\Psi$ 
for $\beta \in \{0.25, 0.5, 1\}$ is illustrated in Fig.~\ref{fig:valPsi}(b).


\begin{figure}[t]
\begin{center}
\begin{tabular}{cc}
\includegraphics[width=0.45\columnwidth]{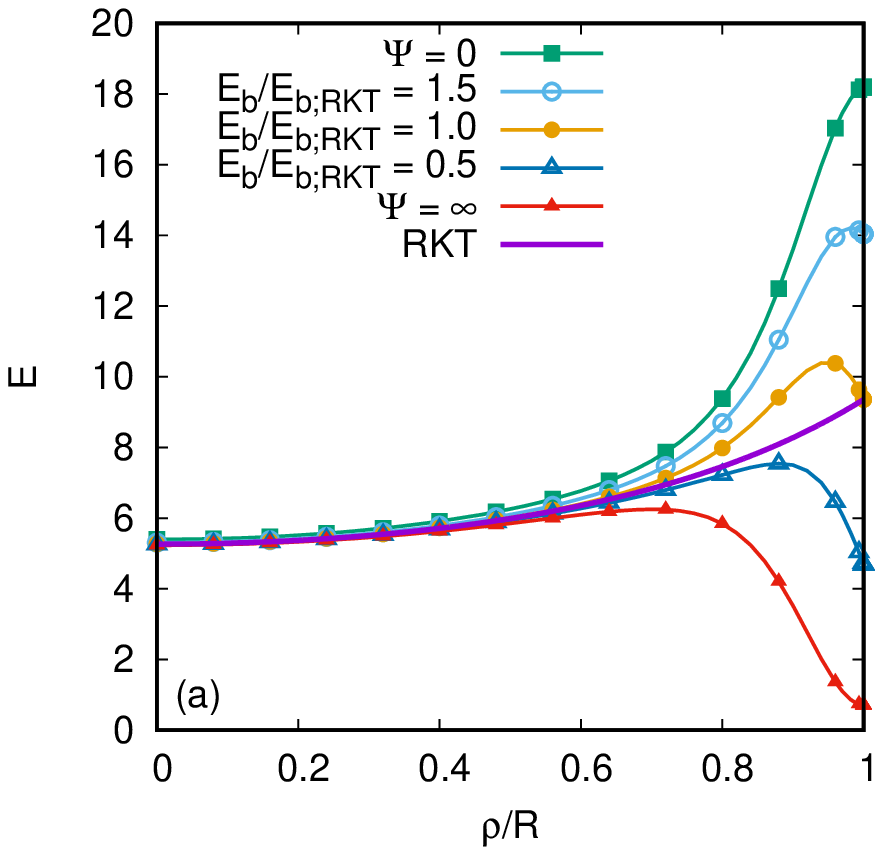}&
\includegraphics[width=0.45\columnwidth]{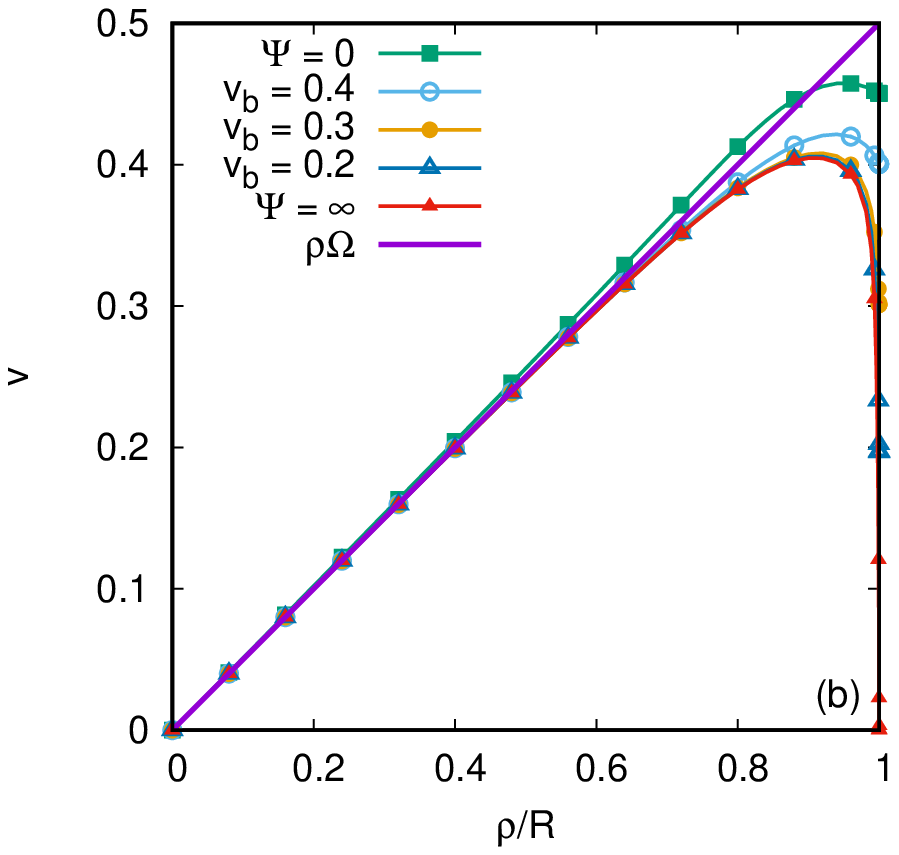}
\end{tabular}
\end{center}
\caption{ 
Profiles of (a) $E$ and (b) $v$ for various values of 
$\Psi$ at $\beta = \Omega = 0.5$ and $R = 1$. 
The QFT results are shown using lines and points, while the 
solid purple line represents the RKT results.
\label{fig:profiles}}
\end{figure}

Finally, we examine the profiles of the energy density and velocity
when $\beta = 0.5$, $\Omega = 0.5$ and $R = 1$. 
Fig.~\ref{fig:profiles}(a) shows that the RBCs 
interpolate between the Dirichlet and von Neumann b.c.s.
In the former case, the energy density exhibits a strong decreasing 
trend in the vicinity of the boundary, as also remarked in Ref.~\cite{duffy03}. 
For the von Neumann b.c.s, the energy density is amplified
next to the boundary, as compared to the 
RKT prediction for a rigidly-rotating Bose-Einstein gas. The velocity 
$v$, shown in Fig.~\ref{fig:profiles}(b), shows small variations with 
respect to $\Psi$. 

\section{Conclusion}\label{sec:conc}

In this paper, a procedure to correlate the parameter $\Psi$ of 
the RBCs for the massless KG field enclosed within a cylinder and the 
boundary values of the rigidly-rotating t.e.v. of the SET operator 
was introduced. 
The restriction $\Psi \ge 0$ was imposed in order to 
eliminate modes with negative co-rotating energy 
$\widetilde{\omega}$ which would otherwise cause t.e.v.s to diverge.
Employing the Landau frame decomposition to obtain the macroscopic 
four-velocity of the state, the velocity on the boundary was shown to 
take values between $0$ (Dirichlet limit) and a maximum value
(von Neumann limit),
which increases towards the value corresponding to rigid rotation
as the temperature is increased. The energy density is 
strongly quenched compared to the relativistic kinetic 
theory prediction for a rigidly-rotating Bose-Einstein gas
in the vicinity of the boundary when the Dirichlet b.c.s 
are employed. By contrast, it is amplified in the case of 
the von Neumann b.c.s. 


\section*{Acknowledgments}
This work was supported by a grant of Ministry of Research and Innovation, CNCS-UEFISCDI,
project number PN-III-P1-1.1-PD-2016-1423, within PNCDI III.


\bibliographystyle{plain}
\bibliography{vambrus}

\end{document}